\documentclass[epj]{svjour}

\usepackage{amsmath,amssymb,amsfonts,graphicx,dcolumn}%, pifont}
\usepackage{epstopdf}
\usepackage{color}
\usepackage{fontenc}

\usepackage{blindtext}

\usepackage{graphicx}
\begin{document}

% Please give the surname of the lead author for the running footer

\title{Deeply virtual Compton scattering off Helium nuclei with positron beams}

\author{Sara Fucini\inst{1} \and Mohammad Hattawy\inst{2}
\and Matteo Rinaldi\inst{1} \and Sergio Scopetta\inst{1}% etc
% \thanks is optional - remove next line if not needed
%\thanks{\emph{Present address:} Insert the address here if needed}%
}                     % Do not remove
%
%\offprints{}          % Insert a name or remove this line
%
\institute{
Dipartimento di Fisica e Geologia, Università degli studi di Perugia, 
and INFN, sezione di Perugia, via A. Pascoli snc, 06123, Perugia, Italy
\and 
Old Dominion University, Norfolk, Virginia 23529, USA.
}
\date{Received: date / Revised version: date}

%TC:break Abstract
%the command above serves to have a word count for the abstract
\abstract{
%\blindtext
Positron initiated deeply virtual Compton scattering (DVCS) off $^4$He and $^3$He nuclei
is described. The way the so-called $d-$term could be obtained from the 
real part of the relevant Compton form factor is summarized, and the importance and novelty of this measurement is discussed. The measurements addressed for $^3$He targets
could be very useful even in a standard unpolarized target setup, measuring beam spin { and beam charge} asymmetries  only.
The unpolarized beam charge asymmetries for DVCS off $^3$He and $^4$He are also estimated, at JLab kinematics and, for $^4$He, also at a configuration typical at the future Electron-Ion Collider.
Incoherent  DVCS processes, in particular the ones with tagging  the internal target by measuring slow recoiling nuclei,  and the unique possibility offered by positron beams for the 
   investigation of Compton form factors of higher twist, are also  briefly 
   addressed. 
    }

\maketitle
  
%TC:break main
%the command above serves to have a word count for the abstract

%\begin{keywords}
%bla | bla | bla | bla
%\end{keywords}

%\begin{corrauthor}
%\texttt{r.henriques{@}ucl.ac.uk}
%r.henriques\at ucl.ac.uk
%\end{corrauthor}

\section*{Introduction}
{In recent years, a growing interest on nuclear deeply virtual Compton scattering (DVCS), i.e., hard photon electroproduction from nuclear targets, has arisen. This is mainly motivated by the possibility to shed light on the European Muon Collaboration (EMC) effect, i.e., the elusive nuclear modifications of the nucleon parton structure (see, e.g., Refs. \cite{Dupre:2015jha,Cloet:2019mql} for recent reports), as well as the possibility to distinguish the so-called coherent and incoherent channels of the DVCS process. This latter feature has been experimentally recently achieved by the CLAS collaboration at JLab using a $^4$He gaseous target \cite{Hattawy:2017woc,Hattawy:2018liu,Dupre:2021uco}, paving the way to a new class of precise measurements at high luminosity facilities. Coherent DVCS takes place when the nucleus recoils elastically, while in the incoherent process the struck proton is detected in the final state. Recently, the measurement of positron initiated DVCS has been experimentally proposed, in particular at JLab with the 12 GeV electron beam \cite{Accardi:2020swt}. In the present paper, we analyze the impact that these measurements may have using $^4$He and $^3$He targets. This is done separately for the coherent and incoherent channels in the next two sections. Some additional remarks on higher twist effects are reported in the fourth section, followed by the conclusions of our investigation}.

 \begin{figure}[tb]
 \centering 
 \includegraphics[width=7.5cm]{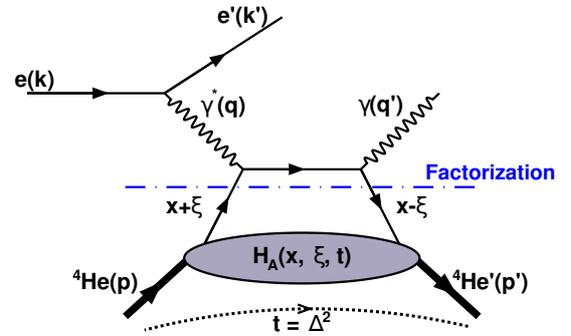}
 \caption{The leading-order, twist-2, handbag approximation to the coherent DVCS process off $^4$He, where the four-vectors of the initial/final electron, photon, and $^4$He nucleus are denoted by $k/k^\prime$, $q/q^\prime$, and $p/p^\prime$, respectively. The quantity $x+\xi$ is the hadron's longitudinal momentum fraction carried by the struck quark, -2$\xi$ is the longitudinal momentum fraction of the momentum transfer $\Delta$ ($= q - q^\prime=p^\prime -p$), and $t$~($=\Delta^2$) is the squared momentum transfer.}
 \label{fig:diags}
 \end{figure}

\section*{Coherent DVCS}
For the sake of concreteness,
to explore the insights on how positron beams could help in this field, 
let us think first to  coherent DVCS off $^4$He,
the only process which has been clearly accessed so far, illustrated, in the handbag
approximation, in Fig.~\ref{fig:diags}.
In DVCS, the relevant non-perturbative information is encoded
in the so called generalized parton distribution functions (GPDs),
giving access, in specific kinematic configurations, to  the location, in the transverse plane of the target, of one parton of given longitudinal momentum (the so-called hadron tomography).  
%As observed for the nucleon target before in this White Paper,
%positrons allow a precise extraction of the Re CFF.
We recall that $^4$He has only one chiral-even  GPD at leading-twist. 
   
{In the coherent DVCS data analysis from the EG6 experiment by the CLAS collaboration}
\cite{Hattawy:2017woc}, the crucial measured observable was the {single beam-spin} asymmetry, $A_{LU}$, which can be extracted from the reaction yields with the two electron helicities ($N^{\pm}$):
\begin{equation}
A_{LU} = \frac{1}{P_{B}} \frac{N^{+} - N^{-}}{N^{+} + N^{-} },
\end{equation}
where $P_{B}$ is the degree of longitudinal polarization of the incident 
electron beam.

\begin{figure}
\centering
\resizebox{0.50\textwidth}{!}{
\includegraphics[angle=270]{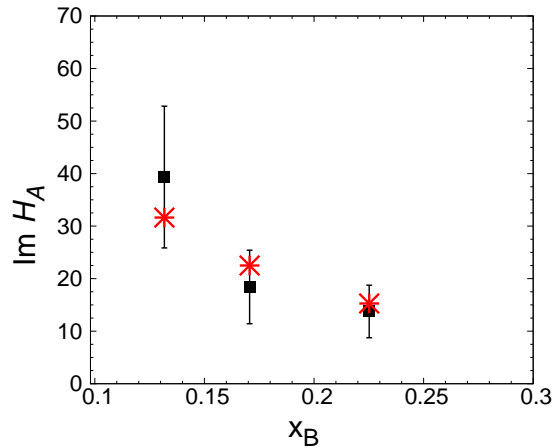}}
\caption{
The imaginary part of the CFF measured in coherent DVCS off $^4$He.
Data from Ref. \cite{Hattawy:2017woc}; calculations (red crosses) from
Ref. \cite{Fucini:2018gso}
}.
\label{uno}
\end{figure}

\begin{figure}%[tbhp]
\centering
%\resizebox{0.75\textwidth}{%
\resizebox{0.50\textwidth}{!}{
\includegraphics[angle=270]{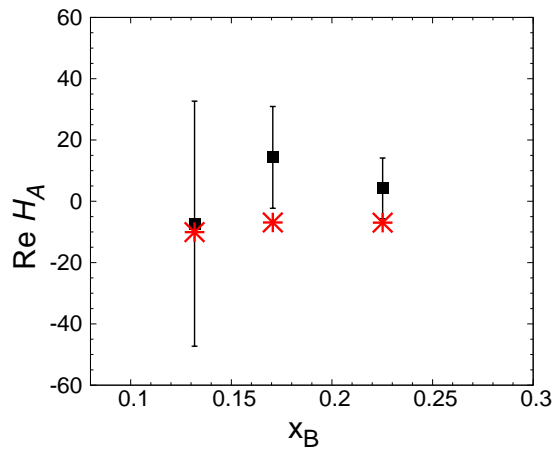}}
\caption{
The real part of the CFF measured in coherent DVCS off $^4$He.
Data from Ref. \cite{Hattawy:2017woc}; calculations (red crosses) from
Ref. \cite{Fucini:2018gso}.
}
\label{due}
\end{figure}

{Within the accessed kinematical phase-space of the EG6 experiment}, the cross section of real photon electroproduction is dominated by the {so-called Bethe-Heitler(BH) process, where the real photon is emitted by the incoming or the outgoing lepton}, while the pure DVCS contribution is very small. However, the DVCS contribution is enhanced in the observables sensitive to the interference term, {\it i.e.} the quantity $A_{LU}$ given above, which depends on the azimuthal angle $\phi$ between the $(e,e^\prime)$ and $(\gamma^*,^4$He$^\prime)$ planes. The asymmetry $A_{LU}$ for a spin-zero target can be approximated at leading-twist as
\begin{equation}
A_{LU}(\phi) = 
\frac{\alpha_{0}(\phi) \, \Im m(\mathcal{H}_{A})} 
{den(\phi)} \, ,
\end{equation}
with:
\begin{eqnarray}
den(\phi) & = & 
\alpha_{1}(\phi) + \alpha_{2}(\phi) \, \Re e(\mathcal{H}_{A}) 
\nonumber
\\
& + & \alpha_{3}(\phi) \, 
\big( \Re e(\mathcal{H}_{A})^{2} + \Im m(\mathcal{H}_{A})^{2} \big)\, .
\label{boh}
\end{eqnarray}
The kinematic factors $\alpha_i$ are known (see, e.g., Ref.  
\cite{Belitsky:2001ns,Belitsky:2008bz}). 
In the experimental analysis,
using the different contributions proportional to
$\sin(\phi)$ and 
$\cos(\phi)$ in Eq. (\ref{boh}),  both the real and 
imaginary parts of the so-called Compton Form Factor (CFF) $\mathcal{H}_{A}$,
$\Re e(\mathcal{H}_{A})$ and $\Im m(\mathcal{H}_{A})$, respectively,
have been extracted by fitting the $A_{LU}(\phi)$ distribution. 
Results of the impulse approximation calculation  
of Ref. \cite{Fucini:2018gso}, are shown together with the data 
of Ref. \cite{Hattawy:2017woc} in Figs. \ref{uno} and \ref{due}.
In the theoretical calculation,
use is made of state-of-the-art ingredients for the description of the nuclear and nucleon structure (older calculations are found in
\cite{Liuti:2005gi,Guzey:2003jh}).
In particular, a convolution formula is obtained where the nuclear effects are governed by a model of a one-body non-diagonal
spectral function based on the Av18 nucleon nucleon interaction \cite{Wiringa:1994wb}
and the Urbana IX three body forces \cite{Pudliner:1997ck} (see Ref. \cite{Fucini:2018gso} for details),
while the nucleon GPDs are modelled following
the Goloskokv-Kroll model 
\cite{Kroll:2012sm,Goloskokov:2008ib,Diehl:2013xca}.
Big statistical errors are seen everywhere in the data but, in particular, $\Re 
e(\mathcal{H}_{A})$ is less precisely extracted
than $\Im m(\mathcal{H}_{A})$, due to the small 
coefficient $\alpha_2$ in Eq. (\ref{boh}).
This fact is easily understood looking at Fig. 3
where it is apparent that the predicted contribution of the real part of the CFF to the symmetry is really small.

Forth-coming data from JLab with 12 GeV electron beams at high luminosity, using also the detector system
developed by the ALERT run-group \cite{Armstrong:2017wfw}, will be affected by much smaller statistical
errors. Together with refined realistic theoretical calculations, in progress for light nuclei \cite{Fucini:2020vpr},
the new data will help 
to unveil a possible exotic behavior of the real and imaginary part of $\mathcal{H}_{A}$, beyond that predicted in a conventional realistic scenario using the Impulse approximation.
Nonetheless, the extracted $\Re e(\mathcal{H}_{A})$ will be always less precise than
$\Im m(\mathcal{H}_{A})$, intrinsically, due to the small coefficient $\alpha_2$
in (\ref{boh}) previously introduced.
A precise knowledge of $\Re e(\mathcal{H}_{A})$ for light nuclei would be instead crucial, as it is 
briefly 
analyzed in what follows, specified initially
for the $^4$He target. Formally one can write,
for the quantities $\Re e(\mathcal{H}_{A})$ and $\Im m \mathcal{H}_{A}$
shown in Figs.~\ref{due} and \ref{uno} respectively \cite{Guidal:2013rya}:
\begin{equation}
\Re e \mathcal{H}_{A} (\xi,t) \equiv 
{\cal P} \int_0^1 dx H_+(x,\xi,t) C_+(x,\xi) \, ,
\label{ReC}
\end{equation}
and
\begin{equation}
  \Im m \mathcal{H}_{A} = -\pi H_+(\xi,\xi,t)  \, ,
\end{equation}
with:
\begin{equation}
    H_+ = H(x,\xi,t)-H(-x,\xi,t) \, ,
\end{equation}
amd
\begin{equation}
    C_+(x,\xi) = \frac{1}{x+\xi}+\frac{1}{x-\xi} \, ,
\end{equation}

with $H(x,\xi,t)$ the chiral-even, leading twist generalized parton distribution (GPD).

Besides, it is also known that $\Re e(\mathcal{H}_{A})$ satisfies a once 
subtracted dispersion relation at fixed $t$ and can be therefore related
to $\Im m \mathcal{H}_{A}$, leading to 
\cite{Anikin:2007yh,Diehl:2007jb,Radyushkin:2011dh,Pasquini:2014vua}
\begin{equation}
\Re e \mathcal{H}_{A} (\xi,t) \equiv
{\cal P} \int_0^1 dx H_+(x,x,t) C_+(x,\xi)
- \delta(t) \, .
\label{disp}
\end{equation}

\begin{figure}{t}
\hspace{-8.5cm}
    \includegraphics[scale=1.1]{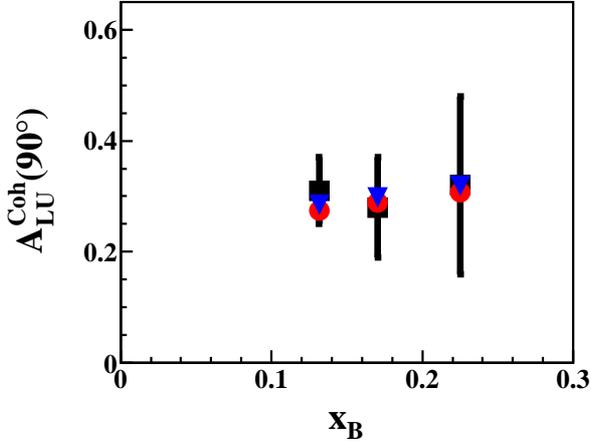}
    \caption{The beam spin asymmetry 
    measured in coherent DVCS off $^4$He,
    for $\phi=90^o$. 
Data from Ref. \cite{Hattawy:2017woc}; calculations (red dots) from
Ref. \cite{Fucini:2018gso}. The result obtained 
    neglecting the real part of the CFF $\mathcal{H}_A$ is also shown (blue triangles)}
    \label{bsa_4he}
\end{figure}

One notices that, in contrast to the convolution integral defining the real 
part of the CFF in Eq. (\ref{ReC}), where the GPD enters for unequal values of its first and second argument, the integrated in the dispersion relation
corresponds to the GPD where its first and second arguments are equal.
The subtraction term $\delta(t)$ can be related to the so-called $d-$term and
accurate measurements, supplemented by precise calculations, would allow therefore to study
this quantity for nuclei, for the first time.
This $d-$term, introduced initially to recover 
the so-called polinomiality property in  double distribution (DD) approaches to GPDs modelling \cite{Polyakov:1999gs},
has been related to the form factor of the QCD energy momentum tensor (see e.g. Ref.  
\cite{Polyakov:2018zvc,Dutrieux:2021nlz}). It encodes information on the distribution of forces 
and pressure between elementary QCD degrees of freedom in the target. For
nuclei, it has been predicted to behave as $A^{7/3}$ in a mean field scheme, 
either in the liquid drop model of nuclear structure \cite{Polyakov:2002yz}
or in the Walecka model \cite{Jung:2014jja}. None of these approaches makes 
much sense for light nuclei, for which accurate realistic calculations are possible. 
Using light nuclei one would therefore explore, at the parton 
level, the onset and evolution of the mean field behavior across the
periodic 
table, from deuteron to $^4$He, whose density and binding are not far
from those of finite nuclei.
\vskip 0.2cm
In this sense, coherent DVCS off $^3$He targets acquire an important role: an intermediate 
behavior is expected between that of the almost unbound deuteron system and 
that of the deeply bound alpha particle. The formal description of coherent DVCS off $^3$He follows that for the proton
\cite{Belitsky:2001ns,Belitsky:2008bz},
a spin one-half target, in terms of CFFs and related GPDs, accessed properly
defining spin dependent asymmetries. Realistic theoretical calculations
are available for GPDs 
\cite{Scopetta:2004kj,Scopetta:2009sn,Rinaldi:2012ft,Rinaldi:2012pj,Rinaldi:2014bba} and are in 
progress for the relevant CFFs, cross sections and asymmetries, representing
an important support to the planning of measurements \cite{Fucini:2020vpr}.
{We remark that in Refs. \cite{Rinaldi:2012ft,Rinaldi:2012pj} it has been shown that the $^3$He chiral-even  GPDs are strongly dominated by those of the neutron, which can be safely extracted by properly taking into account the nuclear effects described in the Impulse Approximation.   }
{One could object that the use of $^3$He, either longitudinally or transversely polarized, represents at the moment a challenge, either with electron or positron beams. Actually beam-charge asymmetries, built using
electron and positron data, would represent, even with unpolarized $^3$He targets and unpolarized beams, a possible access to
$\Re e \mathcal{H}_{A} (\xi,t)$,
as previously described for $^4$He, with the same potential to explore the "d-"term for this relevant light nucleus.}

Positron beams would guarantee this achievement: as a matter of fact, combining data for properly defined asymmetries 
measured 
using electrons and positrons, the role of $\Re e \mathcal{H}_{A}$ would be 
directly accessed. Let us recall how it is possible.

One should notice that, between the quantities appearing in the above
equations and the cross sections defining the generic photo-$e^\pm$ production,
 in the following schematic general expression \cite{Accardi:2020swt}:
\begin{eqnarray}
\sigma^e_{\lambda 0}  & = & \sigma_{BH} + \sigma_{DVCS} + \lambda \tilde \sigma_{DVCS} 
\nonumber
\\
& + & e\sigma_{INT} + e \lambda \tilde \sigma_{INT} \, ,
\label{gen}
\end{eqnarray}
the following relations hold:
\begin{eqnarray}
\sigma_{BH} & \propto & \alpha_1(\phi)\, ,
\nonumber \\
\sigma_{DVCS} & \propto &   \alpha_{3}(\phi) 
\big( \Re e(\mathcal{H}_{A})^{2} + \Im m(\mathcal{H}_{A})^{2} \big) \, ,
\nonumber \\
\sigma_{INT} & \propto & \alpha_{2}(\phi) \, \Re e(\mathcal{H}_{A}) \, ,
\nonumber \\
\tilde \sigma_{INT} & \propto & \alpha_{0}(\phi) \, \Im m(\mathcal{H}_{A}) \, ,
\end{eqnarray}
while $\tilde \sigma_{DVCS} $ is proportional to a term kinematically suppressed at JLab kinematics, dependent on higher twist CFFs. 

From a combined analysis of data taken with polarized electrons and/or 
positrons, one could access all the five cross sections in Eq. (\ref{gen}).
{We stress in particular that, using just unpolarized electrons and positrons, $\Re 
e(\mathcal{H}_{A})$ would be directly accessed, building charge beam asymmetries}.
To expose this fact, in order to extract the real part of CFF, the  beam charge asymmetry (BCA$_A$), for a nucleus with $A$ nucleons, is introduced. By following the same formalism adopted in Ref. \cite{Belitsky:2001ns}, one can define the BCA$_A$ as follows:

\begin{align}
 \mbox{BCA}_A= \dfrac{d \sigma^+-d\sigma^-}{d \sigma^+-d\sigma^-}~,   
\end{align}
where here $d\sigma^\pm$ represents the five times differential unpolarized cross section for  an electron ($-$) or a positron ($+$) beam, respectively. Let us recall that such a cross section for, e.g.,  the process  $e^\pm~ ^3He \rightarrow e^\pm~ ^3He \gamma$ 
\cite{Belitsky:2001ns,Belitsky:2008bz}, reads:

\begin{align}
 \label{cros1}
d\sigma^C\equiv \dfrac{d\sigma^C}{dx_B~dy~dt~d\phi~d \varphi} = 
\dfrac{\alpha^3 x_B 
y}{16 \pi^2 Q^2 \sqrt{1+\varepsilon^2}} \left| \dfrac{T_C}{e^3} \right|^2
\end{align}
where $\varepsilon=2 x_B M_3/Q$, $C=\pm 1$ is the lepton charge,  $M_3$ is the $^3$He mass and $y$ is the lepton energy fraction.
Moreover $\phi$ is the azimuthal angle between the lepton plane and the recoiled nucleus. Finally, 
$T^2_C=|T_{BH}|^2+|T_{DVCS}|^2+I_C$. Let us remark that only
the interference term depends on the 
beam charge \cite{Belitsky:2001ns}. Therefore:

\begin{align}
\label{diff1}
 d\sigma^+-d\sigma^- = \dfrac{2 \alpha^3 x_B 
y}{16 \pi^2 Q^2 \sqrt{1+\varepsilon^2}}  \dfrac{I}{e^6} 
\end{align}
where $I=I_C/C$. Within this formalism:

\begin{align}
 \label{inter}
I &= \dfrac{e^6}{x_B y^3 t P_1(\phi) P_2(\phi)}
\\
\nonumber
&\times \Big\{c_0^I+\sum_1^3 
\big[c_n^I \cos(n \phi)+s_n^I \sin(n \phi) \big]  \Big\}.
\end{align}
These harmonic coefficients  depend  on the $^3$He CFFs. In particular, if unpolarized targets and beams are considered, $s_n^I$ do not contribute to the BCA$_3$. The functions $P_1$ and $P_2$ \cite{Belitsky:2001ns} cancel out in the evaluation of BCA$_3$. On the other hand:

\begin{align}
\label{sum1}
 d\sigma^++d\sigma^- = \dfrac{2 \alpha^3 x_B 
y}{16 \pi^2 Q^2 \sqrt{1+\varepsilon^2}}  
\dfrac{|T_{DVCS}|^2+|T_{BH}|^2}{e^3}~. 
\end{align}
 We remind that, when terms of the order $t/Q^2$ can be neglected, BCA$_3$ is directly related to the real part of CFFs. As a matter of facts, in this kinematic conditions,
 
\begin{align}
\mbox{BCA}_3(\phi) \sim \dfrac{e^6}{x_B y^3 \Delta^2 P_1(\phi) P_2(\phi)} \dfrac{c_1^I 
\cos(\phi)}{|T_{BH}|^2}~.
\end{align}
We remind that, since
$c_3^I$ is related to the gluon distribution, in this initial analysis such a contribution is also neglected, together with 
$c_0^I$ and $c_2^I$. Finally $c_1^I \propto Re~c_{unp}^I $ where 
$c_{unp}^I=F_1 \mathcal{H}+x_B/(2-x_B)(F_1+F_2)\mathcal{\tilde H}-\Delta^2/(4 M^2)F_2 \mathcal{E}$ (for details see Ref. \cite{Belitsky:2001ns}). We recall that $F_1$ and $F_2$ are the Dirac and Pauli form factors respectively and $\mathcal{F}=\mathcal{H,E, \tilde H}$ is the CFF. Finally:

\begin{align}
\mbox{BCA}_3(\phi) =\dfrac{x_B (1+\varepsilon^2)^2}{y} \dfrac{c_1^I \cos(\phi)}{ 
c_0^{BH}+c_1^{BH} \cos(\phi)}~.
\label{bca3}
\end{align}

\begin{figure}
\hskip -0.2cm
    \includegraphics[scale=0.48]{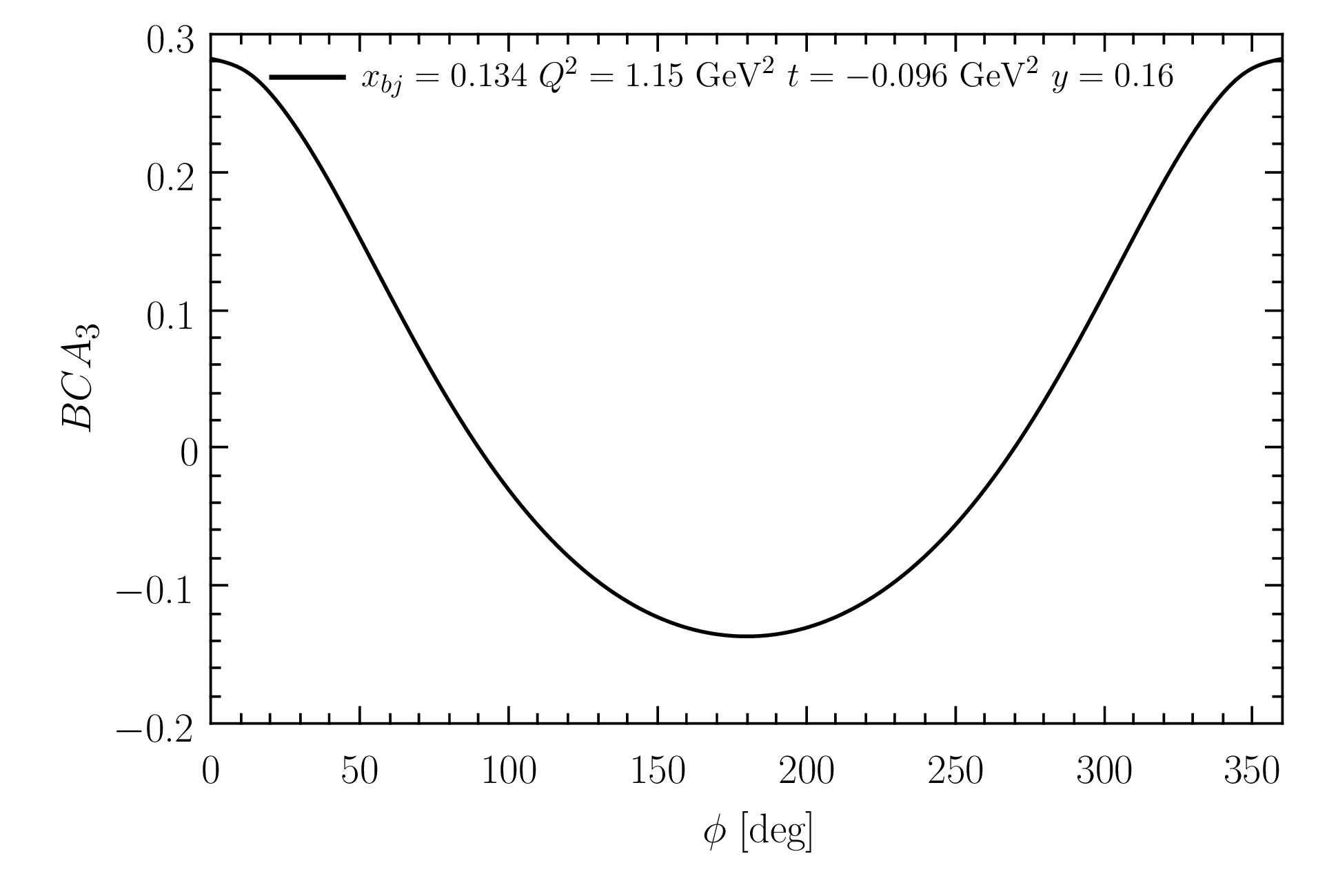}
    \hskip -0.2cm
        \includegraphics[scale=0.48]{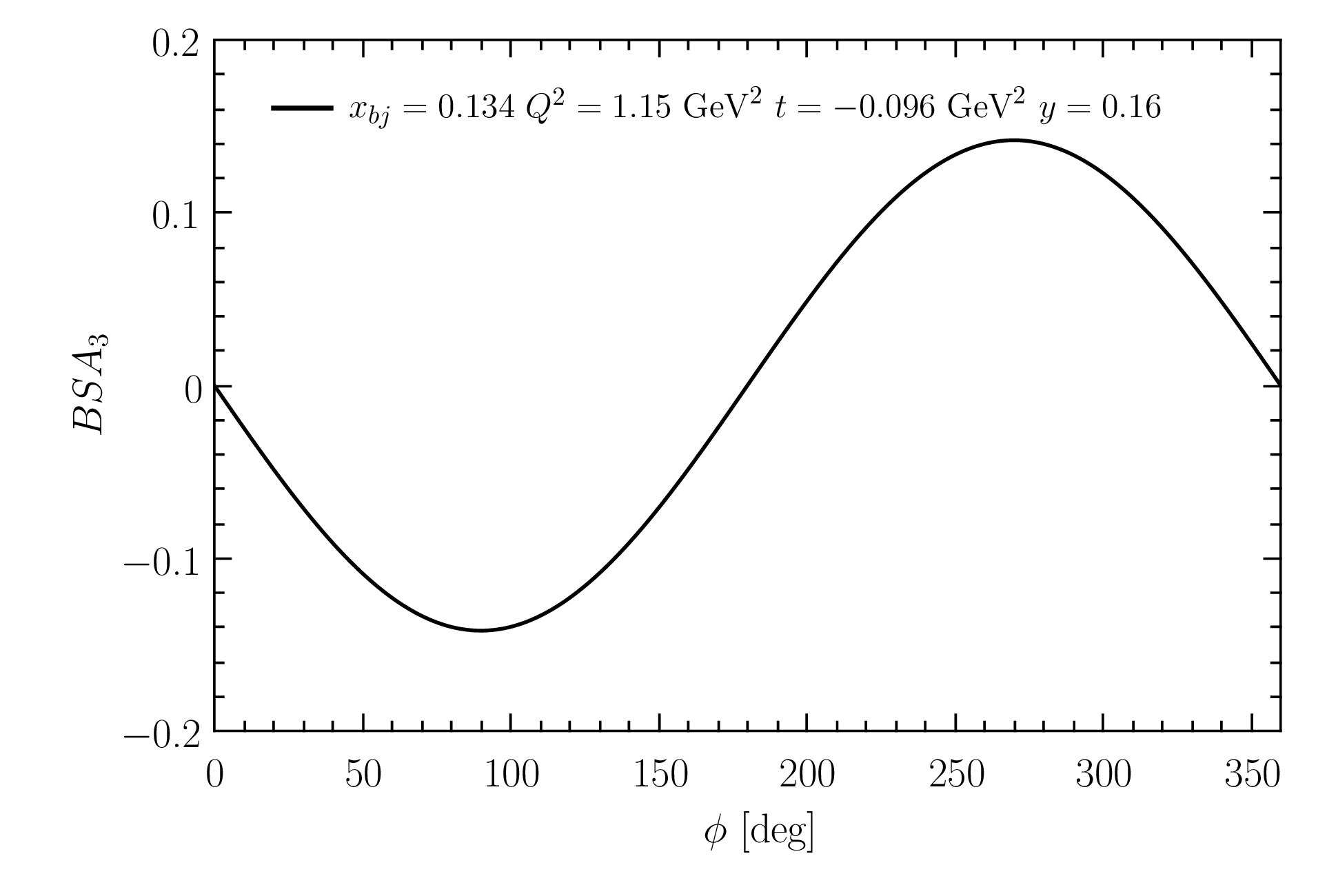}
    \caption{Upper panel: The $^3$He beam charge asymmetry Eq. (\ref{bca3}) for a specific kinematic configuration, typical at JLab. Lower panel: the same but for the $^3$He beam spin asymmetry
    Eq. (\ref{bsa3}).}
    \label{bca_3he}
\end{figure}

Numerical predictions for the $^3$He BCA,
evaluated in a kinenatical range typical at JLab, are shown for the first time in the upper panel of Fig. \ref{bca_3he}. Here the $^3$He GPDs have been calculated by following the line of Refs. \cite{Rinaldi:2012ft,Rinaldi:2012pj}, i.e. by means of the off-forward spin dependent spectral function calculated  from the $^3$He wave function corresponding to the AV18 nuclear potential \cite{Pace:2001cm,Kievsky:1996gz,Kievsky:1997bg}. The nucleonic GPDs, necessary as an input to the calculations, have been chosen among the phenomenological parametrizations of Refs. \cite{Kroll:2012sm,Goloskokov:2008ib,Diehl:2013xca}.
In addition one can introduce the nuclear beam spin asymmetry (BSA):

\begin{align}
 \mbox{BSA}_A= \dfrac{d \sigma^{\uparrow}-d\sigma^{\downarrow}}{d \sigma^{\uparrow}+d\sigma^{\downarrow}}~.   
\end{align}
By using the previously discussed strategy and approximations, one can relate the above quantity to the imaginary part of CFF. For a 1/2 spin target like the $^3$He one gets:

\begin{align}
 \mbox{BSA}_3 \sim \pm  \dfrac{x_{Bj}}{y} \dfrac{s_{1}^I}{c_{0}^{BH}}   \sin(\phi)~,
 \label{bsa3}
\end{align}
where now $s^I_1 \propto~ Im~c^I_{unp}$.

\vspace{0.2cm}

\begin{figure}
    \centering
    \includegraphics[scale=0.6]{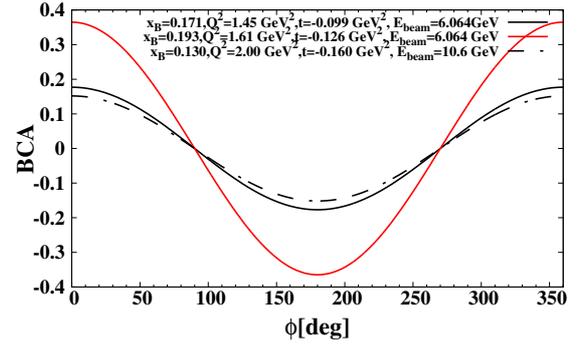}
    \caption{The $^4$He beam charge asymmetry shown at different kinematic points, typical at JLab.}
    \label{bca_4he_EIC}
\end{figure}

\begin{figure}
    \centering
    \includegraphics[scale=0.6]{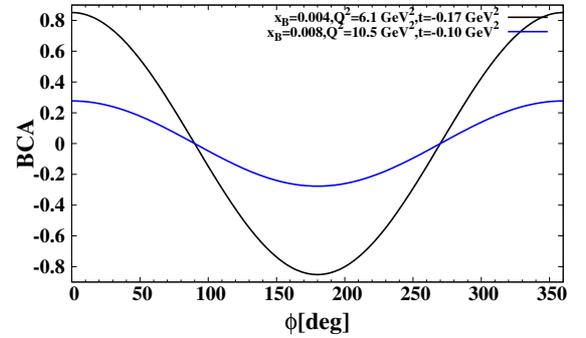}
    \caption{The $^4$He beam charge asymmetry shown at different EIC kinematics. The energy configuration corresponding to these plots is $E_{electron}\times E_{nucleus}=(18x110)$ GeV.}
    \label{bca_4he}
\end{figure}

The formalism above discussed for the beam charge asymmetry of the $^3$He nucleus can be easily extended to the $^4$He nucleus. While the beam spin asymmetry for a longitudinally polarized electron beam off an unpolarized target has already demonstrated to be a key observable to access the the imaginary part of the CFF, the BCA is mostly useful to extract information for the real part Eq. \eqref{disp}. Looking at Eq.~\eqref{bca3}, the numerator cleanly encapsulates the real part of CFF linked to the chiral even GPD of the $H_q^A$. Conversely to the $^3$He case, the $^4$He spinless nature allows one the access to only this GPDs without the contamination of other GPDs.
We show here our first estimates of the BCA for $^4$He.
We make use, for the expression of the GPD $H_q^A$ of the impulse approximation model presented in Ref. \cite{Fucini:2018gso} where, as already reported, a semi-realistic model for the nuclear part, accounting for a realistic momentum dependence obtained with realistic NN potential (Av18) and three-body forces, was adopted. While the non diagonal momentum and the energy dependencies of the spectral function is just modeled,  a confirmation of the capability  of our model to describe conventional nuclear effects is at hand. As an example of this fact, one can check Figs. \ref{uno}-\ref{due}. In passing by, we remind that the twist-two CFF of the $^4$He nucleus receives the contribution of the GPD $H$ and $E$ of the bound nucleons in the form $H_q^{N/A}=\sqrt{(1-\xi^2)}\bigg[ H_q^N - \frac{\xi^2}{1-\xi^2}E_q^N\bigg]$.
As noted in Ref. \cite{Fucini:2020vpr}, the contribution of the GPD $E_q^N$ is practically negligible at the kinematics probed at JLab due to the smallness of the skweness (an even greater kinematical suppression for such a factor will be observed in the typical kinematic ranges
expected at the Electron Ion Collider (EIC) \cite{AbdulKhalek:2021gbh}). For the GPDs of the nucleons, we made use of the same GK models used for $^3$He, i.e. the ones described in Refs. \cite{Goloskokov:2006hr,Goloskokov:2008ib}.
In Fig. \ref{bca_4he}, a sizable asymmetry is observed for $^4$He plotted ad various kinematics. Two of these are the same already probed in Ref. \cite{Hattawy:2017woc}, while another one  is the same foreseen for the free proton
using positron beams (see. Ref. \cite{Burkert:2021rxz}). As a reference, we show in Fig. \ref{bca_4he_EIC} the same quantity for one of the kinematic points  envisioned at the EIC for a positron beam at 18 GeV colliding with an helium-4 beam with an energy of 110 GeV. The kinematic points plotted is the present paper have been obtained as pseudo-data generated with the software \texttt{TOPEG}.\\
While in Ref. \cite{AbdulKhalek:2021gbh}  the possibility to reach the first tomographic view of the nucleus is shown, here the big asymmetry found in some of the kinematical ranges investigated, allowing one an easier extraction of information from the observable,  confirms the prominent role that the EIC together with the upgraded JLab will play with the aim to reach a deeper comprehension of the innermost structure of hadronic targets and hadronic matter.

\begin{figure}
\centering
\includegraphics[width=7.5cm]{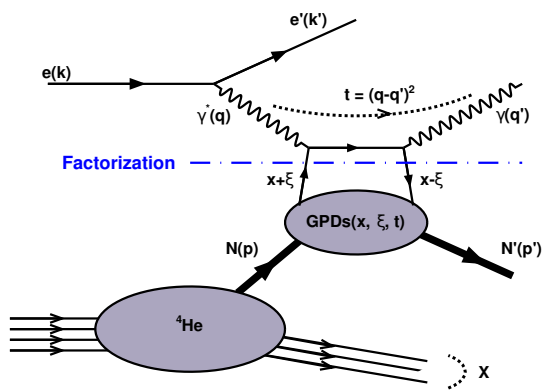}
\caption{The leading-order, twist-2, handbag approximation to incoherent DVCS off $^4$He.
See Fig.~\ref{fig:diags} for the definition of the variables.} 
\label{fig:diags2}
\end{figure}

\section*{Incoherent DVCS}

A subject aside is represented by incoherent DVCS off He nuclei, i.e.,
the process where, in the final state, the struck proton is detected, its CFFs accessed, its GPDs, in principle, extracted and, ultimately, its tomography obtained. Fig.~\ref{fig:diags2} illustrates the leading-twist
handbag diagram for the incoherent DVCS process off $^4$He. Such measurements would provide a pictorial representation of the realization of the EMC effect and a great progress towards the understanding of its dynamical origin. As already stressed, this channel has been successfully isolated by the EG6 experiment of the CLAS collaboration at JLab \cite{Hattawy:2018liu} and a first glimpse at the parton structure of the bound proton in the transverse coordinate space is therefore at hand. Recent impulse approximation calculations, using again a
description based on the Av18 interaction
for the nuclear effects and, for the nucleon part, 
the GK model and the one proposed in Ref \cite{Mezrag:2013mya},
evaluated using the PARTON tool \cite{Berthou:2015oaw}, 
have been presented
in Refs \cite{Fucini:2019xlc,Fucini:2020lxi}.
The theoretical description
of the recent data with conventional realistic ingredients is rather encouraging.
The program at JLab 12 includes an improvement of the accuracy of these measurements, in particular, for the first time in DVCS, tagging the struck nucleon detecting
slow recoiling nuclei, using the detector
developed by the ALERT run group \cite{Armstrong:2017zcm}.
This would allow to keep possible final state interactions, relevant in principle in this channel, under control.
Measurements performed with 
electron and positron beams  would allow for example the measurement of the $d-$term for the bound nucleon, either proton in $^3$He (tagging 2H from DVCS on $^3$He)  or in $^4$He (tagging $^3$H from 
DVCS on $^4$He) or neutron in $^4$He (tagging $^3$He from DVCS on $^4$He). Modifications 
of the $d-$term of the nucleon in the nuclear medium, studied e.g. in
Ref.
\cite{Jung:2014jja}, would be at hand, as well as a glimpse at the structure of the neutron
in the transverse plane, complementary to that obtained with deuteron targets.

\section*{Beyond a chiral even GPDs description of DVCS on $^4$He}

{As a last argument,
we quickly note that, from the measurement of 
beam spin asymmetries built using 
cross sections measured with polarized electrons and positrons
in coherent DVCS off $^4$He, the 
terms $\tilde 
\sigma_{DVCS}$
and $\tilde 
\sigma_{INT}$,
appearing in Eq. \ref{gen},
could be independently accessed.
This would allow, for example for $^4$He,
to study the other leading twist CFF of a spinless target,
the so called  gluon transversity GPD $H_T$, giving a corresponding
name to the CFF ${\cal{H}}_T$, appearing in
$\tilde  \sigma_{DVCS}$. In Ref. 
\cite{Belitsky:2008bz}, it is shown how
the contribution of ${\cal{H}}_T$ to the cross section
occurs through an interference between twist-two
and effective twist-three CFFs. A first glimpse at this complicated interrelation
would be obtained for a spin-less target, in particular for a nuclear target,
for the first time.
As for any other gluon-sensitive observable, data for the cross section
$\tilde  \sigma_{DVCS}$ would be a perfect tool to study gluon dynamics in nuclei. For example, a comparison between the above observable and calculations performed in an Impulse Approximation scheme, where the relevant nuclear degrees of freedom
are colorless nucleons and mesons, would
have the potential to expose possible exotic gluon dynamics in nuclei.
This would be a pretty new possibility, complementary to that planned at JLab
with 12 GeV, using exclusive vector meson electroproduction off
$^4$He \cite{Armstrong:2017zcm}.
Such an interersting behavior would be very hardly seen using electrons only,
due to the strong kinematical suppression of $\tilde  \sigma_{DVCS}$ with respect to
the other contributions in Eq. (\ref{gen}).
}

\section*{Conclusions}
{
The unique possibilities offered by the use of positron beams in DVCS off three- and four-body nuclear systems have been reviewed. Summarizing, we can conclude that the main advantages will be:

\begin{itemize}
    \item in coherent DVCS off $^3$He and $^4$He, using polarized electrons and unpolarized positrons,
    the real part of the chiral even unpolarized CFFs would be measured with a precision comparable to that of their imaginary part, providing a tool for the study of the so called $d$-term. In turn,  the distribution of pressure and forces between the partons in nuclei, a new way to look at the nuclear medium modifications of nucleon structure, could be investigated; 
    \item in incoherent DVCS off $^3$He and $^4$He, possibly tagging slow recoiling nuclear systems, the same programme could be run for the bound proton and neutron;
    \item using polarized $^3$He targets, a more complicated setup for the moment, spin dependent and parton helicity flip CFFs would be accessed for the first time for a nucleus, in both their real and imaginary parts. 
    {Moreover, in this case these quantities would be dominated by the neutron contributions and, in turn,  an  extraction of the neutron CFFs would be feasible};
    \item coherent DVCS off $^4$He, initiated with polarized positrons, would allow a first analysis of nuclear chiral odd CFFs and GPDs, with higher twist contamination suitable to tentatively explore gluon dynamics in nuclei.
\end{itemize}

A programme of nuclear measurements with positron beams would represent therefore an exciting complement to the experiments planned with nucleon targets, and to those planned with nuclear targets and electron beams.
In this paper, we have shown for the first time realistic Impulse Approximation predictions for spin-independent charge asymmetries,
at JLab kinematics for $^3$He and $^4$He and, in this last case, also in one of the typical kinematic setups foreseen at the EIC. In an extended forthcoming investigation these calculations will be extended to the incoherent channels and to other spin dependent asymmetries. 
}

%\section*{Results}

%Just for kicks here's a citation \cite{Gustafsson2016}. And a %reference to a supplement \cref{note:Note1}. And %\nameref{note:Note1}.
%\Blindtext

%\Blindtext

%Figure \ref{fig:computerNo} shows an example of how to insert a column-wide figure. To insert a figure wider than one column, please use the \verb|\begin{figure*}...\end{figure*}| environment. Figures wider than one column should be sized to 11.4 cm or 17.8 cm wide. Use \verb|\begin{SCfigure*}...\end{SCfigure*}| for a wide figure with side captions.

%\Blindtext \Blindtext \Blindtext

%\section*{Conclusions}

%$blablaba \ref{fig:computerNo} 
%\blindtext

%\subsection*{Blabla} 
%\blindtext

%\section*{Conclusions}

%blablaba \ref{fig:computerNo}
%\blindtext

%\begin{acknowledgements}
%\blindtext
%\end{acknowledgements}

\newpage

%\section*{Bibliography}
\bibliographystyle{unsrt}
\bibliography{MyWPCont}

%% You can use these special %TC: tags to ignore certain parts of the text.
%TC:ignore
%the command above ignores this section for word count
\onecolumn
\newpage

%\section*{Word Counts}
%This section is \textit{not} included in the word count. 
%\subsection*{Notes on Nature Methods Brief Communication}
%\begin{itemize}
%\item Abstract: 3 sentences, 70 words.
%\item Main text: 3 pages, 2 figures, 1000-1500 words, more figures possible if under 3 pages
%\end{itemize}

%\subsection*{Statistics on word count}
%\detailtexcount
%\newpage

 %%%%%%%%%%%%%%%%%%%%%%%%%%%%%
%% Supplementary Information %
%%%%%%%%%%%%%%%%%%%%%%%%%%%%%%
%\captionsetup*{format=largeformat}
%\section{Full list of authors and affiliations} \label{note:Note1} 

%S.~Fucini$^a$, M.~Hattawy$^b$, M.~Rinaldi$^a$, S.~Scopetta$^a$  

%\begin{center}

%{\it
%$^a${Dipartimento di Fisica e Geologia, Università degli studi di Perugia, and 
%   INFN, sezione di Perugia, \\
%   via A. Pascoli snc, 06123, Perugia, Italy}

% $^b${Old Dominion University, Norfolk, Virginia 23529, USA.}
% }
%\end{center}

%%TC:endignore
%%the command above ignores this section for word count

\end{document}